\documentclass[twocolumn, traditabstract]{aa} 
\usepackage{graphicx}
\usepackage{natbib}
\bibpunct{(}{)}{;}{a}{}{,} 
\usepackage{amsmath} 
\usepackage{subeqnarray}
\usepackage{layouts}
\usepackage{color}
    \definecolor{Blue}{rgb}{0.0,0.0,1.0}
    \definecolor{Red}{rgb}{1.0,0.0,0.0}
    \definecolor{Green}{rgb}{0.0,1.0,0.0}

\definecolor{battleship}{rgb}{0.77,0.75,0.81}

\newcommand{\be}{\begin{equation}}
\newcommand{\ee}{\end{equation}}
\newcommand{\bea}{\begin{eqnarray}}
\newcommand{\eea}{\end{eqnarray}}

\begin{document}
\title{Escape, capture, and levitation of matter in Eddington outbursts}
\author{Adam Stahl\inst{1} 
  \and W{\l}odek Klu\'zniak\inst{2}
  \and Maciej Wielgus\inst{3} 
  \and Marek Abramowicz\inst{1,2}}

\institute{Physics Department, Gothenburg University, SE-412-96
  G{\"o}teborg, Sweden\\ \email{gusstaad@student.gu.se}
  \\ \email{marek.abramowicz@physics.gu.se}
  \and Nicolaus Copernicus Astronomical Center, ul. Bartycka 18,
  PL-00-716 Warszawa, Poland \\ \email{wlodek@camk.edu.pl}
  \and Institute of Micromechanics and Photonics, ul. {\'s}w
  A. Boboli 8, PL-02-525, Warszawa, Poland  \\ \email{maciek.wielgus@gmail.com}
}
\date{Received April 1, 2013; accepted May 26, 2013}
\abstract{
{\it Context:} An impulsive increase in luminosity by one half or more
of the Eddington value will lead to ejection of all optically thin
plasma from Keplerian orbits around the radiating star, if gravity
is Newtonian and the Poynting-Robertson drag is neglected.
Radiation drag may bring some particles down to the stellar surface.
{\color{Green}}{ On the other hand, general relativistic
 calculations show that gravity may be balanced by
a sufficiently intense radiation field at a certain distance from the star.
\par\noindent
{\it Aims:} We investigate the motion of test particles around highly
luminous stars to determine conditions under which plasma may be ejected
from the system.}
\par\noindent
{\it Results:}
In Einstein's gravity, if the outburst is close to the Eddington luminosity,
all test particles orbiting outside an ``escape sphere''
will be ejected from the system, while all others will be captured
from their orbits onto the surface of another sphere,
which is well above the stellar surface, and may even be outside
the escape sphere, depending on the value of luminosity.
Radiation drag will bring all the captured particles to
rest on this ``Eddington capture sphere,'' where they will remain
suspended in an equilibrium state
as long as the local flux of radiation does not change and remains
at the effective Eddington value.}

\maketitle

\section{Introduction}
It is widely believed that in luminous compact sources, such as the
accreting neutron stars in low-mass X-ray binaries, Poynting-Robertson
drag will tend to increase the accretion rate as the luminosity
of the source increases.
{\color{Red}}
{Not intending to disagree that in steady sources
radiation drag may enhance the accretion rate,
here}
 we wish to point out that a {\sl sudden increase} in luminosity
{may}
 in fact have the opposite effect {\color{Red}}{in optically thin plasma}:
atoms (or ions) subjected to increased
radiation pressure will tend to move out, away from the stellar surface,
Poynting-Robertson drag notwithstanding.

In Newtonian dynamics, the effects of increased radiation pressure
are straightforward to compute. A simple application of orbital mechanics
reveals that an increase in luminosity by one half of the Eddington
value\footnote{More precisely, an impulsive increase in luminosity by one half
of the difference between the most recent steady value and the Eddington
luminosity is sufficient to unbind a non-relativistic
 test particle \citep{pkorona}.}
is sufficient to eject optically thin plasma from its Keplerian orbits
at any initial distance from the star.

In Einstein's general relativity (GR) this remains qualitatively true at
a large enough distance from the star---in the optically 
thin regime a sudden increase in luminosity of sufficient
magnitude will unbind the orbiting fluid.
However, closer to the star, competing effects of gravity and radiation
pressure may lead to very nonintuitive behavior, particularly when
combined with the effects of radiation drag. We show in an
exact GR calculation that if the luminosity at infinity is close 
to the Eddington value, optically thin matter in the vicinity
of a compact star is neither ejected from the system nor accreted
onto the star. Instead, it will collapse onto the surface of an imaginary
sphere, whose radius is a function of the stellar radius and the luminosity
(Eq.~[\ref{Eddington-radius}]). 
Rather than orbit the star, fluid on this Eddington
Capture Sphere remains at rest in a state of equilibrium
\citep{Stahl}, suspended
above the stellar radius, in what is to all intents and purposes
a state of levitation.

\section{The Eddington capture sphere}

In Einstein's general relativity, the redshifted luminosity $L(r)$
of an isotropic star
decreases according to  
\begin{equation}
L(r)=L_\infty/(1-R_{\rm Schw}/r).
\label{lumr}
\end{equation}
%
The {\em effective} Eddington luminosity at $r$, i.e., the luminosity
at which radiation pressure balances gravity, is
\begin{equation}\label{effective}
L_{\rm eff}(r)=~L_{\rm Edd}(1-R_{\rm Schw}/r)^{-1/2}.
\end{equation}
%
For a given luminosity $L_\infty$,
 a static balance between the radiation pressure force
and gravitational pull is then achieved at the radius at which
$L(r)=L_{\rm eff}(r)$ \citep{Phinney}.
The formula for this unique radius, $r_{\rm Edd}(L_\infty)$,
can be written in the form
\begin{equation}
\label{Eddington-radius}
\frac{r_{\rm Edd}}{R_{\rm Schw}} = {\left[1 - 
\left(\frac{L_\infty}{L_{\rm Edd}} \right)^2\right]^{-1}}\ .
\end{equation}
We use standard Schwarzschild coordinates, with
the Schwarzschild radius  $R_{\rm Schw}=2GM/c^2$, and
$L_\infty/L_{\rm Edd}$ the stellar luminosity at infinity in units of the
Eddington value, $L_{\rm Edd} = 4 \pi G M m_p c/ \sigma$, where $\sigma$
is the Thomson cross-section. 
Strictly speaking the formula is true for 
an optically thin shell of hydrogen
plasma, but similar formulae hold for other compositions of the fluid,
as well as other photon momentum absorption processes.
\cite{Abramowicz} showed that $r_{\rm Edd}$ is a position of
{\color{Green}}{ stable} equilibrium
for test particles (with proton mass, and Thomson cross-section for
photon momentum absorption) moving radially in the combined gravitational
and radiation fields of a~spherical, non-rotating,
isotropically radiating compact star.

It is now known that the sphere at $r_{\rm Edd}$ is also a position of equilibrium
with respect to non-radial perturbations, and that in fact particles
from a wide class of trajectories will be captured on this sphere, because
of radiation drag \cite[see e.g.,][{\color{Green}}{ for work in the Schwarzschild metric}]{Bini,Oh2011,Stahl}.
Hence, we refer to it as the {\it Eddington capture sphere (ECS)}.
{\color{Green}}{ For Kerr metric results, see e.g., \cite{Oh2010, Semerak}.}

\section{Equations of motion}

We carry out all calculations in the Schwarzschild metric using the
standard Schwarzschild spherical coordinates, and extend the treatment
of \cite{Abramowicz} to non-radial motion; we use their stress-energy
tensor of radiation $T^{(i)(k)}$, which was calculated in the stationary observer
tetrad assuming isotropic emission from the stellar surface.
{\color{Green}}{
An explicit form of the stress-energy tensor is given in the Appendix,
along with a summary of the derivation of the equations of motion.}
Spherical symmetry of the problem assures that
for any set of initial conditions the trajectory of a test particle is
confined to a single plane, which we take to be the equatorial plane
$\theta=\pi/2$.

From this point on we take $G=1=c$, and scale all radii with $M$
(i.e., one half of the Schwarzschild radius).
The proper time is given by the Schwarzschild metric restricted to the
equatorial plane
\begin{equation}
d\tau^2 = \left(1-\frac{2}{r}\right)\,dt^2 
 - \left(1-\frac{2}{r}\right)^{-1}\,dr^2 - r^2 d\phi^2.
\label{metric}
\end{equation}
We introduce $B = 1 - 2/r$, and denote the four-velocity by $u^i$, so
 $u^t=d t/d\tau$. The angular frequency observed at infinity
is $\Omega= (d\phi/d\tau)/u^t$,
{\color{Green}}{ when discussing circular orbits
we use its Keplerian value $\Omega_{K}(r)= 1/r^{3/2}\equiv v_{K}(r)/r$.
In the figures, we  find it convenient to use
$v^\phi\equiv r\Omega$, scaled with the coordinate speed of light,
 at those (initial)  moments when the radial
velocity is zero, and $v^r\equiv dr/dt$ when the azimuthal velocity
is zero.}
The stellar radius is denoted by $R$.

{We make the standard assumption of efficient electromagnetic coupling
between protons and electrons and, as is usual in the discussion
of Eddington luminosity, take the mass of the test particle to be equal
to that of the proton. In so doing, we neglect possible effects of
charge separation discussed by \cite{Walker88}.}
{\color{Red}}
{The effects of  Poynting-Robertson
drag in steady disk accretion were considered by \cite{Walker}.
We consider a time-dependent problem. On the assumption
}
that the test particle is absorbing momentum (from the
radiation field)  with the Thomson
cross-section in its rest frame, thus suffering
a rest-frame force of $(\sigma/c)\times$~flux, the particle trajectory
is described by two coupled, second order differential equations:
\begin{eqnarray}
\frac{d^2 r}{d \tau^2}
\!&=&\! \frac{k}{ \pi I(R) R^2}\ \left(BT^{(r)(t)}u^t -
\left[ T^{(r)(r)}+ \varepsilon \right]\frac{d r}{d \tau} \right)+
\nonumber \\
&&+ \left(r-3\right)\left( \frac{d\phi}{d \tau} \right)^2 - \frac{1}{r^2},
\label{system-1}\\
\frac{d^2 \phi}{d \tau^2} \!&=&\!
 - \frac{d \phi}{d \tau} \left(\frac{k}{\pi I(R) R^2}
 \left[T^{(\phi)(\phi)} + \varepsilon \right]
 + \frac{2}{r} \frac{dr}{d\tau}\right),
\label{system-2}
\end{eqnarray}
with
\begin{eqnarray} \nonumber
\varepsilon \!&=&\!
B T^{(t)(t)}(u^t)^2 + B^{-1}T^{(r)(r)}
\left( \frac{dr}{d\tau}\right)^2 +\\
&& + r^2 T^{(\phi)(\phi)} \left(\frac{d \phi}{d \tau} \right)^2 -
2T^{(r)(t)}u^t \frac{dr}{d\tau}.
\label{varepsilon}
\end{eqnarray}
The parameter $k$ is the stellar surface luminosity in Eddington units,
\begin{equation}
k = \frac{L(R)}{L_{\rm Edd}} = 4\pi^2 R^2 \frac{I(R)}{L_{\rm Edd}}
 =\frac{L_\infty}{L_{\rm Edd}}\left(1-\frac{2}{R}\right)^{-1}.
\label{klum}
\end{equation}
Because $BT^{(r)(t)}=\pi I(R)(1-2/R)(R^2/r^2)$,
Eqs. (\ref{system-1}), (\ref{system-2}) allow a static
solution \mbox{$r(\tau)=r_{\rm Edd}$}
(Eq.~[\ref{Eddington-radius}]), provided that

\begin{equation}
\left(1 - 2/R \right)^{1/2}\le L_\infty/L_{\rm Edd}<1.
\label{limits}
\end{equation}
By integrating Eqs.~(\ref{system-1}) and ({\ref{system-2}),
 we are able to find the particle orbits, such as the
ones illustrated in Fig.~\ref{Fig:orbits} 

All numerical results presented below were obtained with the
Dormand-Prince method, which is a fourth-order accuracy, adaptive
Runge-Kutta type integrator.

\begin{figure}
\centering
\includegraphics[width=0.48\textwidth]{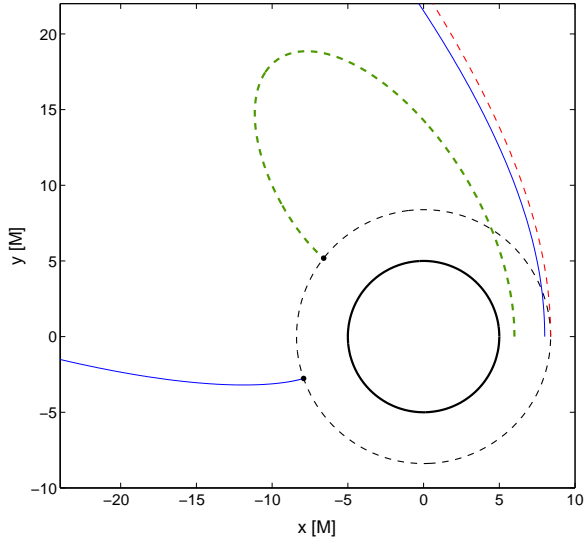}
\caption{Trajectories of particles leaving their circular orbits
following an impulsive increase in luminosity  from zero to
that value ($k=1.45$ for $R=5$) at which the escape sphere
and the Eddington capture sphere coincide.
Particles initially in circular orbits at
(or outside) the escape sphere leave the system
on outward bound trajectories (thin dashed red line).
Those from initial orbits inside the escape sphere,
turn back and are captured by the ECS, on which they remain at rest
in their final position (thick dots).
}
\label{Fig:orbits}
\end{figure}

\section{Escape from the vicinity of a luminous star}

\begin{figure}
\centering
\includegraphics[width=0.48\textwidth]{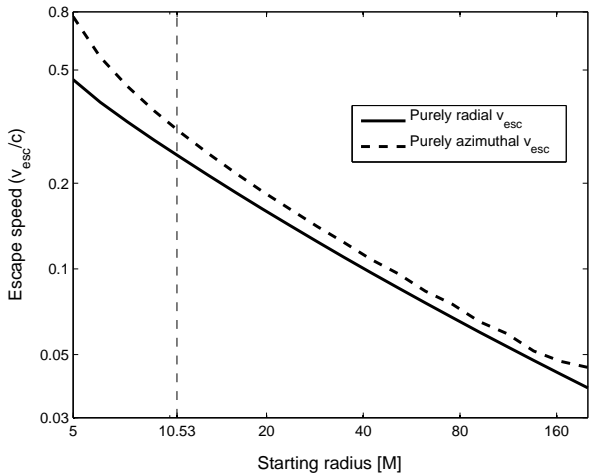}
\caption{Escape speeds normalized to the speed of light,
as a function of the starting radius, for
$R=5, L_\infty=0.9\,L_{\rm Edd}$. Solid line corresponds to radial motion.
The Eddington capture sphere is at
the position marked with a vertical dashed line.}
\label{Fig:asradius}
\end{figure}
\begin{figure}
\centering
\includegraphics[width=0.48\textwidth]{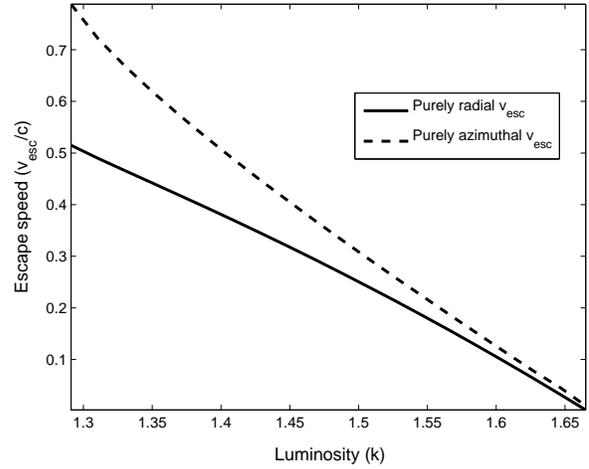}
\caption{Escape speeds (normalized to the speed of light)
from the Eddington capture sphere vs. luminosity for $R=5$.
The~whole range of
luminosities is shown, for which there is an ECS. }
\label{Fig:aslum}
\end{figure}

In the absence of radiation, it is easy enough
to determine whether or not a test particle will leave the system.
A particle is bound or
unbound depending on its angular momentum and energy. An unbound
test particle  is certain to escape to infinity, if only it is outward
going, i.e., if ${\color{Green}}{u}^r(0)\ge 0$,
 while incoming particles may, or may not, be
captured by the star depending on their impact parameter.

The presence of radiation drag complicates
matters, as energy and angular momentum are no longer constants of the
motion.  One needs to compute the actual trajectory for given initial
conditions to find out whether or not the particle leaves the system.
In this paper we only consider trajectories with ${\color{Green}}{u}^r(0)\ge 0$.
For examples of inward bound trajectories see, e.g., \cite{Stahl}.

We have integrated the equations of motion to find trajectories that are not
bounded\footnote{Trajectories that are not bounded cannot be enclosed
in any sphere, however large.}
and found the minimum (``escape'') velocities required for escape from
the system. These depend on the initial velocity vector, with the highest
value of the escape speed corresponding to
 initial velocities tangent to a circle
concentric with the spherical star (``purely azimuthal'' initial motion),
and the lowest for radial motion ($v^\phi=0$). 
 These  two  limiting cases are shown  in
Fig.~\ref{Fig:asradius} as a function of the radius when
the stellar radius and luminosity are $R=5$, and $L_\infty=0.9\,L_{\rm Edd}$
or $k=1.5$.

Technically, the criterion we used to decide whether or not a test particle
escapes to infinity was to integrate the equations of motion until the particle
turned around $({\color{Green}}{u}^r<0$), or conversely,
 reached $r=1500 R_{\rm Schw}$.
In the latter
case, motion was always radial to a high accuracy, and radiation
drag negligible. The particle was deemed to be on a bounded trajectory
 if its (relativistic) specific
energy at $r=1500 R_{\rm Schw}$ was less than unity.
If the specific energy was $\ge1$,
 the particle was classified as escaping the system.

Except at the very highest luminosities, it is rather hard for
particles to escape from the ECS. Fig.~\ref{Fig:aslum} 
shows the values of escape velocities from the {\color{Green}}{ECS},
as a function of the luminosity parameter $k$
of Eq.~(\ref{klum}). 

In the limit of
$L_\infty\rightarrow L_{\rm Edd}$, the ECS becomes infinitely large,
$r_{\rm Edd}\rightarrow \infty$,
and the escape velocity goes to zero. 
In the opposite (lower) limit
of Eq.~(\ref{limits}), $r_{\rm Edd}= R$.
In this case, the escape speed from the stellar
surface,
which now coincides with the ECS,
exceeds one half of the speed of light (Fig.~\ref{Fig:surface})
 for all $R\le 5$. This is  not an unusually high value, if one recalls
\citep{kw}
that in the Schwarzschild metric the highest Keplerian orbital speed 
 is $c/2$ (in the marginally stable orbit, a.k.a. the ISCO, at $r=6$), 
however, at first sight it may seem surprising
that the escape speed remains so high even in the presence
of the effective Eddington luminosity of radiation, whose pressure
balances gravity, Eq.~(\ref{effective}).
As a reminder we note that in this context the
escape speed has a somewhat different meaning than usually.
It is not the speed necessary for the particle to be unbound, i.e.,
its energy to be larger than its rest mass.
Rather, it is that speed which is sufficient to ``overcome''
radiation drag, so that the orbit is not bounded,
i.e., a speed sufficient for the particle to escape
 to infinity in spite of continual loss
of angular momentum to the radiation field.

\begin{figure}
\centering
\includegraphics[width=0.48\textwidth]{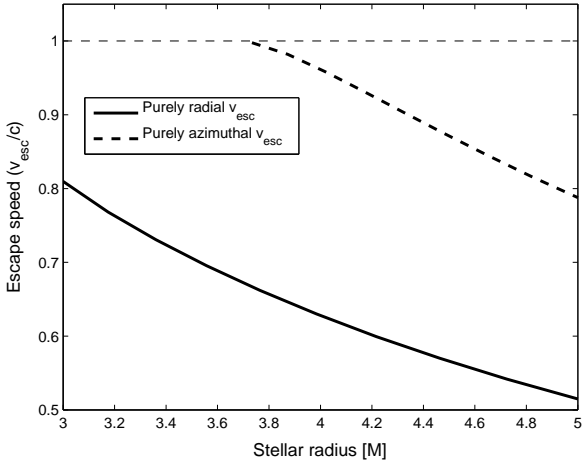}
\caption{Escape speeds from the Eddington capture sphere
when it coincides with the 
stellar surface, as a function of the stellar radius,
 for purely radial and azimuthal initial velocities.}
\label{Fig:surface}
\end{figure}

\begin{figure}
\centering
\includegraphics[width=0.48\textwidth]{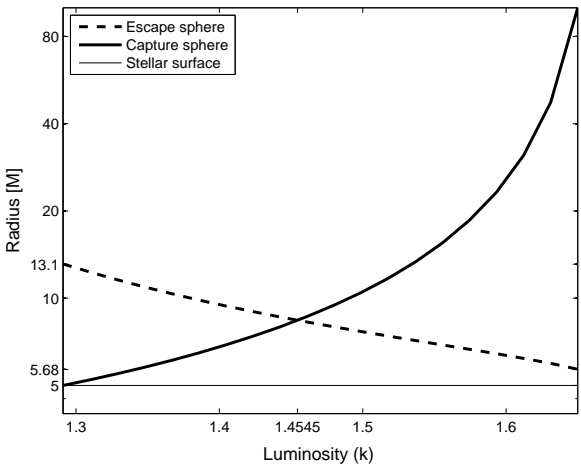}
\caption{Radii of the escape sphere and the Eddington capture sphere
 as a function of
the stellar luminosity ($R=5$). The two spheres coincide at the critical
value of $k=1.45$, or $L_\infty=0.873\, L_{\rm Edd}$, corresponding
to $r_{\rm Edd} =8.40$.
}
\label{Fig:radii}
\end{figure}

\begin{figure}
\centering
\includegraphics[width=0.48\textwidth]{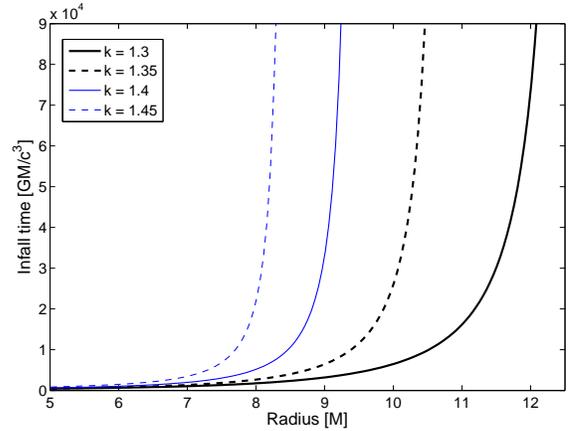}
\caption{Infall time onto the Eddington capture sphere
 from various initial radii within the
escape sphere. Note that the tick marks on the vertical scale are
separated by $10^4 GM/c^3$}
\label{Fig:infall}
\end{figure}

\section{Impulsive change in luminosity}
The question we are addressing in this paper is whether or not an
optically thin corona will be ejected from a system undergoing
an impulsive increase in luminosity to the effective Eddington value,
and what becomes of the particles which are not ejected.
{\color{Green}}{ Astrophysically, such a situation may be encountered,
e.g., in X-ray bursters (see \S~\ref{discuss}).}
In Newtonian mechanics and gravity, this
question was answered by \cite{pkorona}, who found that in
the absence of radiation drag all particles will be ejected
if the luminosity increases above a critical value,
equal to $(L+L_{\rm Edd})/2$, where $L$ was the luminosity prior to the increase.
To solve the problem in full GR with radiation drag included, we integrate
Eqs.~(\ref{system-1}) and ({\ref{system-2}).

Consider the trajectory of a test particle in a circular orbit at $r$
about a star whose luminosity will undergo an impulsive change from zero
to $L(R)$. Thus, the initial conditions correspond to a particle with
$v^r=0$, and $v^\phi=v_K$ appropriate for that circular orbit,
 $v_K$ being the Keplerian {\color{Green}}{speed defined following
 Eq.~\ref{metric}}. 
 We begin
the integration at that moment in which the luminosity at $r$ jumps
from zero to $L(r)$ of Eq.~(\ref{lumr}), and find that 
{\color{Green}}{this initial} $v^\phi$ is higher
than the azimuthal escape speed for all $r$ greater than some critical value,
which is a function of $R$ and $L(R)$, while it is lower
 for all radii below that critical value.
 Accordingly, we
introduce the \emph{escape sphere} dividing space into an outer
region, from which all test particles will be hurled out to infinity
by the impulsive change in luminosity, and the inner region, between
the stellar surface and the escape sphere, from which the particles do
not escape the system
{\color{Green}}{ (c.f. Fig.~\ref{Fig:radii})}. 

The fate of particles within the escape sphere is rather interesting.
Because of radiation drag, for luminosities below the lower limit of
Eq.~(\ref{limits}), the particles would have ended up on the stellar
surface. However, for the luminosities considered here  they all come
to rest on the ECS.  This is illustrated in
Fig.~\ref{Fig:orbits}, which shows an escaping trajectory of a
particle previously in circular orbit on what is now the escape
sphere, as well as bounded trajectories from circular orbits of lower
radii. 

As it turns out, the escape sphere may be inside or outside the ECS.
Thus, for a high enough luminosity, all particles initially
outside the escape sphere escape to infinity, while those inside
it migrate outside to settle on the ECS.
Fig.~\ref{Fig:radii} shows the radii of the escape sphere and of the
capture sphere, as a function of the luminosity parameter. The two
coincide for a critical value of the luminosity parameter,
which we find to be $k_{\rm crit}=1.45$ for $R=5$, the orbits shown in
Fig.~\ref{Fig:orbits} correspond to this case.

For particles that are initially well within the escape sphere, the migration
time to the \color{Green}}{ECS} is rather short. This is illustrated
in Fig.~\ref{Fig:infall}. For a $2M_\odot$ star the unit  of time is
$2GM_\odot/c^3\approx10^{-5}\,$s. Thus, all times shown in the figure
correspond to a fraction of a second, and this is how long the high luminosity
has to be sustained for the particles to collect on the ECS.


\section{Discussion}
\label{discuss}
{\color{Red}}{ While our computations were
  performed for test particles, the main results should also hold for
  plasma, in which internal dissipation of energy may occur. Clearly,
  the particles that eventually settle on the ECS
  lose all their kinetic energy (and angular momentum), and it makes
  little difference whether some of it is first redistributed to other
  particles.  

As the radiation front of increased flux travels at the
  speed of light, all plasma particles are accelerated and the
  outbound particles do not run into other particles departing from
  larger radii.  The particles that are on unbounded trajectories very
  quickly approach radial motion, because of radiation drag.  Very
  little plasma compression and dissipation is likely to occur in this
  situation.  In any case, the corona is assumed to be radiatively
  inefficient, so any dissipation of energy does not lead to binding
  plasma that is unbound---the total energy of the plasma remains
  positive.}
  However, there are two serious limitations of the
current work, which should be overcome in future calculations. We have
assumed spherical symmetry, while the most luminous, and interesting,
astronomical sources accrete in a disk-like geometry.  Further, in the
microscopic domain we have neglected the transfer of momentum between
the scattered (or re-emitted) photons and the test particle---only the
{\color{Red}}{momentum transferred by the absorbed}
 photons was included
in the calculation. Thus the radius of the escape sphere is likely to
have been underestimated. On the other hand, the position of the
 \color{Green}}{ECS} is insensitive to this assumption, particles
at rest on the ECS radiate isotropically, and the only momentum
transfer is from the absorption process included in our calculation.

The future will tell whether the theoretical finding presented here
have any application to X-ray bursts, which are spherically symmetric
to a high degree, and how they affect the estimates
of neutron star radius, as in, e.g., \cite{Damen}. 
{\color{Red}}{As the precise value of the stellar radius and mass is not known,
it is hard at present to translate the observed flux from the neutron star
to a precise value of the luminosity $k$ parameter.
Should a direct observational determination
of $k$ turn out to be possible, a value for the stellar radius would
immediately follow (cf. Eq.~\ref{klum}).}
{\color{Green}}{}{However, we note that X-ray bursters are thought to attain
Eddington luminosity, e.g.,  \cite{Lewin}, and a recent burst in
the neutron star 2S 0918-549 has exhibited an interesting achromatic
oscillation, which has been interpreted as an accretion disk response to
super-Eddington fluxes in an X-ray burst \citep{in't}.
 \cite{Wielgus} point out that,
in principle, a luminosity oscillation may be related to the appearance
and disappearance of the ECS.}
Very interesting
luminosity variations were also reported in the Z sources,
or rather, in the Z phase of bright LMXBs \citep{Lin}, and in some
microquasars---of course, these are precisely the sources where
spherical symmetry is broken, and more work is needed before any
reliable conclusions can be reached on the relevance to these sources
of the phenomena discussed here.

{\color{Green}}{}{Perhaps it is worth noting that super-Eddington luminosity may
 also be present among some accreting black holes, notably ULXs
 \citep{Feng,Gladstone}, but also in supermassive black holes,
 \citep[e.g.,][]{Kollmeier,Nobuta}, and microquasars, such as  
V4641 Sgr in outburst \citep{Revnivtsev}.}


\section{Conclusions}

We have shown in a GR calculation including radiation drag
that a sudden change in luminosity from zero to about
the (effective) Eddington value leads to ejection of all test particles
outside a certain sphere (the escape sphere). 
In Newtonian physics all particles outside the stellar surface are either
ejected, or not, depending on the luminosity jump \citep{pkorona},
when radiation drag is neglected. In Einstein's gravity, the 
redshifted luminosity drops with radius, and this effect, combined with
radiation drag, introduces a radius dependence to the problem, thus
allowing some particle to be left behind in the system.
At the high final luminosities considered here, the non-escaping particles
are not accreted by the star. Instead, they 
{\color{Red}}{lose all kinetic energy and angular momentum and }
{\color{Green}}{ migrate to the Eddington capture sphere,
 typically within a fraction of a second, where they remain at rest}
 as long as the local flux
of radiation corresponds to the effective Eddington value.

We expect our work to be applicable to the optically thin coronae of
variable X-ray sources.  
Future work will show how the position of the escape sphere depends
on the initial luminosity of the system {\color{Green}}{ \citep{Mishra}}.

\section{Acknowledgments}

We thank Saul Rappaport and Wenfei Yu for inspiring discussions.
In addition, Saul Rappaport's help was invaluable in validating
our codes.
Research supported in part by Polish NCN grants
UMO-2011/01/B/ST9/05439, and N N203 511238,
and the Czech grant MSM
4781305903, as well as by a Swedish VR grant.

\section{{Appendix}}
Here, we present an abbreviated derivation of Eqs.~(\ref{system-1}), (\ref{system-2}).
The acceleration of the particle is in general given by 
\begin{equation}
a^{i}=u^{k}\nabla_{k}u^{i}=\frac{d^{2}x^{i}}{ds^{2}}+\Gamma_{jk}^{i}u^{j}u^{k}
 =\frac{1}{M}\frac{d}{d\tau}(u^{i})
  +\Gamma_{jk}^{i}u^{j}u^{k}\ ,\label{eq:genRelAcc}
\end{equation}
where $ds^2=M^2d\tau^2$.
Taking the force of radiation to be
\begin{equation}
c^2ma^{i}=\frac{\sigma}{c}F^{i}\ ,\label{eq:newtons2}
\end{equation}
we obtain the equation of motion
\begin{equation}
\frac{d}{d\tau}(u^{i})=Ma^{i}-M\Gamma_{jk}^{i}u^{j}u^{k}
 =\frac{M\sigma}{mc^{3}}F^{i}-M\Gamma_{jk}^{i}u^{j}u^{k}\ .
\label{eq:generalEOM}
\end{equation}
We determine the radiation flux $F^{i}$ from the expression 
\begin{equation}
F^{i}=h_{\ j}^{i}T^{jk}u_{k}\ ,
\end{equation}
where $h_{\, j}^{i}=\delta_{\, j}^{i}-u^{i}u_{j}$ is the projection
tensor, and $T^{jk}$ is the stress-energy tensor for the radiation.
Thus, the flux is 
\begin{equation}
F^{i}= T^{ik}u_{k}-u^{i}\epsilon\ ,
\label{eq:generalFlux}
\end{equation}
with
\begin{align}
\epsilon &
 \equiv T^{tt}u_{t}u_{t}+2T^{tr}u_{t}u_{r}+T^{rr}u_{r}u_{r}+T^{\phi\phi}u_{\phi}u_{\phi}\ .
\end{align}
Expressing $\varepsilon\equiv\epsilon/[{\pi I(r)]}$ 
in tetrad components, we obtain Eq.~(\ref{varepsilon}).
We take the tetrad components of the stress-energy tensor
from \cite{Abramowicz}:
\begin{align}
T^{(t)(t)} & =2\pi I(r)\left(1-\cos\alpha\right)\\
T^{(t)(r)} & =\pi I(r)\left(\sin^{2}\alpha\right)\\
T^{(r)(r)} & ={\textstyle \frac{2}{3}}\pi I(r)\left(1-\cos^{3}\alpha\right)\\
T^{(\phi)(\phi)} & 
 ={\textstyle \frac{1}{3}}\pi I(r)\left(\cos^{3}\alpha-3\cos\alpha+2\right)\ ,
\end{align}
with all other components except $T^{(\theta)(\theta)}$ vanishing.
The $T^{(\theta)(\theta)}$-component, when it appears, will however
always be multiplied by $u^{\theta}=0$, so we may ignore this component
for the present problem. In the equations above, $\alpha=\alpha(r)$
is the viewing angle and $I(r)$ is the intensity of radiation.
 In our dimensionless coordinates, they are given by 
\begin{align}
\sin\alpha & =\frac{R}{r}\sqrt{\frac{1-2/r}{1-2/R}}\,,\\
I(r) & =I(R)\left(\frac{1-2/R}{1-2/r}\right)^{2}.\label{eq:intensity}
\end{align}


\end{document}